# Microstrip-Line Characterization of Artificial Dielectric Materials for MRI Applications

Haiwei Chen, Huan Wang, Xiaotong Zhang, *Senior Member, IEEE*, Yang Gao, *Member, IEEE*

***Abstract*— Artificial dielectric materials (ADMs) offer a flexible means of shaping radiofrequency fields in magnetic resonance imaging (MRI). However, ADM designs based on plane-wave characterization may not directly predict their effects on RF fields in MRI. We propose a travelling-wave microstrip-line (MSL) approach to approximate the spatially nonuniform excitation experienced by dielectric pads near MRI RF coils and to guide ADM characterization and tuning against a reference dielectric. Both materials are tested under identical MSL measurement conditions, and the ADM geometry is adjusted until its reflection minimum occurs at the same frequency as that of the reference. Simulations and bench measurements show that the tuned ADM more closely matches the reference dielectric's reflection-minimum frequency than a plane-wave-based design. MRI simulations show that the tuned ADM produces a transmit magnetic-field distribution similar to that of the reference pad. Phantom images acquired at 7 T also show similar image-intensity patterns. These results demonstrate the utility of MSL response matching for guiding ADM design in the tested configuration. The proposed approach provides a practical bench-top step for assessing and tuning manufactured ADMs before MRI evaluation.**

***Index Terms*—Artificial dielectric materials, dielectric characterization, high-permittivity materials, magnetic resonance imaging (MRI), microstrip lines.**

## I. Introduction

Magnetic Resonance Imaging (MRI) provides excellent soft-tissue contrast and spatial resolution. Ultra-high-field (UHF) systems operating at 7 T and above offer increased signal-to-noise ratio and opportunities for higher-resolution imaging [1], [2], [3], [4]. However, spatial variations in the transmit RF magnetic field, $B_1^+=(B_x+iB_y)/2$, can produce nonuniform spin excitation, while RF power deposition, quantified by the specific absorption rate (SAR), constrains imaging performance and patient safety [5].

High-permittivity materials can modify the RF field distribution and power deposition [6]-[14]. Their practical implementation may nevertheless be constrained by material availability, fabrication requirements, and cost. Artificial dielectric materials (ADMs) provide an alternative by using engineered structures to reproduce selected electromagnetic properties of high-permittivity dielectrics [14]-[18]. Depending on their implementation, ADMs can offer fabrication flexibility and reduced cost while reproducing the effects of reference dielectric pads on the $B_1^+$ field [19]. Their design and characterization commonly employ effective parameters derived under plane-wave excitation [17], [20].

A variety of microwave techniques are available for dielectric characterization. Transmission and reflection methods, including the Nicolson–Ross–Weir (NRW) approach, retrieve complex constitutive parameters from scattering measurements using models of wave propagation in defined measurement configurations [21], [22]. Resonant techniques, such as dielectric-loaded resonators, provide sensitive characterization at selected frequencies through changes in resonance frequency and quality factor [23]. Open-ended coaxial probe techniques enable broadband, non-destructive permittivity measurements of liquids and biological tissues using admittance-based models of the reflection response at the probe aperture [24], [25]. Planar microwave sensors based on transmission lines and substrate-integrated waveguide (SIW) structures also infer permittivity from changes in scattering parameters or resonant features [26].

These techniques provide established tools for material characterization under defined measurement conditions. For finite, structured ADMs, however, the measured effective response can depend on the excitation polarization and spatial field distribution, as well as the sample geometry. An ADM designed to match a reference dielectric under one excitation condition may produce a different RF field distribution when placed near an MRI coil. For ADM-based dielectric pads, a complementary bench-top method for assessing and tuning this equivalence before MRI evaluation would facilitate practical design and fabrication.

In this study, we introduce a microstrip-line (MSL)-based method for characterizing and tuning the effective electromagnetic response of ADMs for 7 T MRI. The method compares the reflection response of an ADM with that of a reference dielectric in the same configuration under a selected modal excitation. The frequency of the reflection minimum serves as a practical matching criterion associated with the characteristic impedance and electrical length of the loaded line.

This work was supported in part by National Natural Science Foundation of China (52307256, 52277232, 52293424), the Fundamental Research Funds for the Central Universities (XJSJ23009). (*Corresponding author: Yang Gao, gaoyang.ziint@gmail.com*)

Haiwei Chen and Yang Gao are with Hangzhou Institute of Technology, Xidian University, 311200 Hangzhou, China and the Sir Peter Mansfield Imaging Centre, School of Physics and Astronomy, University of Nottingham, NG7 2RD Nottingham, United Kingdom.

Huan Wang is with Hangzhou Institute of Technology, Xidian University, 311200 Hangzhou, China.

Xiaotong Zhang is with the College of Electrical Engineering, Zhejiang University, Hangzhou 310000, China.

Simulations and bench measurements are used to evaluate this approach, followed by MRI simulations and phantom imaging experiments to assess whether the response matching translates to a similar effect in the MRI environment.

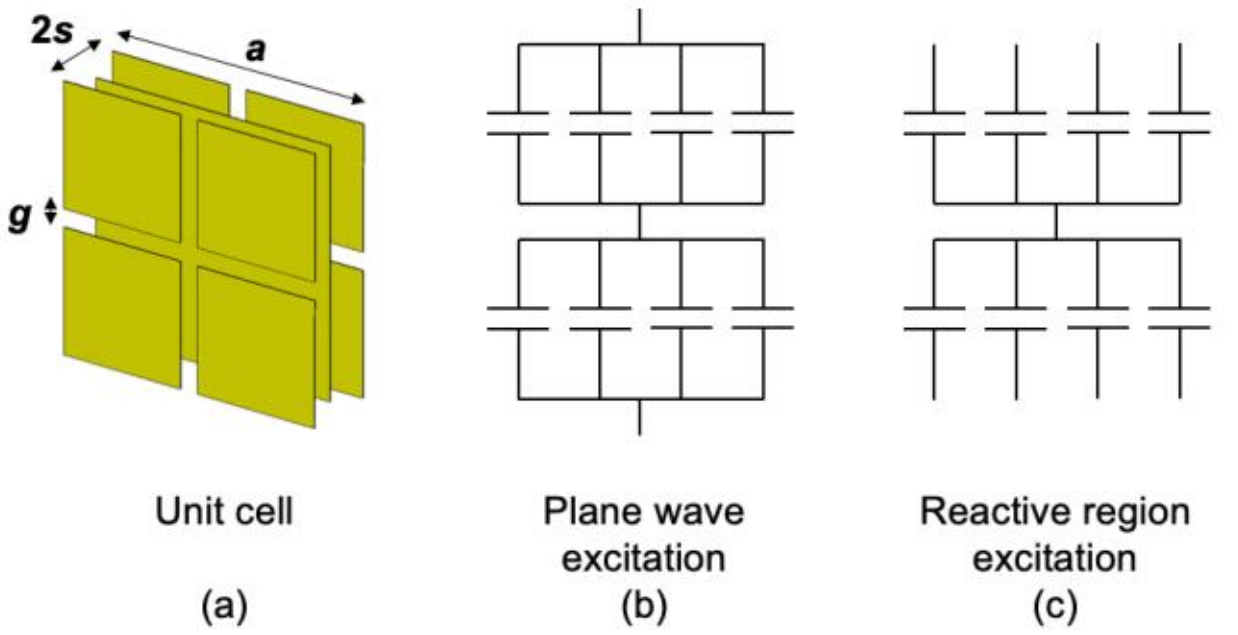

**Fig. 1. Unit Cell Structure and Equivalent Circuits of the ADM:** (a) Schematic diagram of the selected ADM unit cell structure. (b) Equivalent circuit diagram of the ADM under plane wave excitation. (c) Equivalent circuit diagram of the ADM under reactive region excitation. This figure illustrates the fundamental structure and theoretical models of the ADM, highlighting the differences in electrical behavior under different excitation conditions.

## II. Methodology

### *A. ADM Response Under Different Excitation Conditions*

The ADM used in this study follows the structure proposed by Vorobyev et al. [19] and consists of stacked capacitive grids fabricated using printed-circuit boards. Each unit cell comprises two metal patch layers and two insulating layers, with adjacent patch layers laterally offset by half the structural period, as illustrated in Fig. 1(a). This structure was selected because of its demonstrated ability to reproduce the effects of high-permittivity dielectric pads in MRI. Figs. 1(b) and 1(c) present schematic equivalent circuits illustrating the electrical response under plane-wave and spatially nonuniform excitation, respectively [27].

Effective-permittivity estimates based on plane-wave reflection and transmission provide a useful starting point for ADM design [17], [20]. However, a finite ADM placed near an MRI coil experiences a spatially nonuniform field with a different distribution of electric-field components. Its response may therefore differ from that predicted under plane-wave excitation. We investigated this excitation dependence by comparing the surface-current and electric-field distributions within the same ADM structure under plane-wave and birdcage-coil excitation.

A birdcage coil was selected as a representative MRI transmit resonator [28], [29], [30], [31]. Fig. 2 shows the simulation configurations and the corresponding surface-current distributions on the ADM. Fig. 3 compares the electric-field distributions under the two excitation conditions, including the field in the surrounding space and the $E_x$ and $E_z$ components within the ADM. These comparisons were used to examine how the excitation conditions affect the field distribution and induced current paths in the finite structure.

These comparisons motivate the MSL approach described below, in which the ADM and a reference dielectric are compared under the same controlled excitation. Whether the resulting response matching translates to similar behavior in an MRI coil is subsequently evaluated through MRI simulations and phantom experiments.

### *B. Microstrip-Line Characterization and Tuning*

Motivated by the excitation-dependent response described above, we use a travelling-wave microstrip line (MSL) to compare and tune the effective electromagnetic response of an ADM relative to a reference dielectric. The material under test serves as the substrate of the MSL, providing a controlled, spatially nonuniform excitation.

The MSL is terminated with a 50 Ω load to reduce end reflections (Fig. 4(a)). Its input reflection coefficient, S11, depends on the characteristic impedance and electrical length of the loaded line, as well as losses and termination conditions. Replacing the reference dielectric with an ADM changes these properties and may shift the frequency of the reflection minimum. Here, this minimum is interpreted as an input-impedance matching condition, which is related to the characteristic impedance and electrical length. With the remaining MSL geometry and measurement conditions held constant, its frequency is used as a practical criterion for tuning the ADM response to that of the reference dielectric.

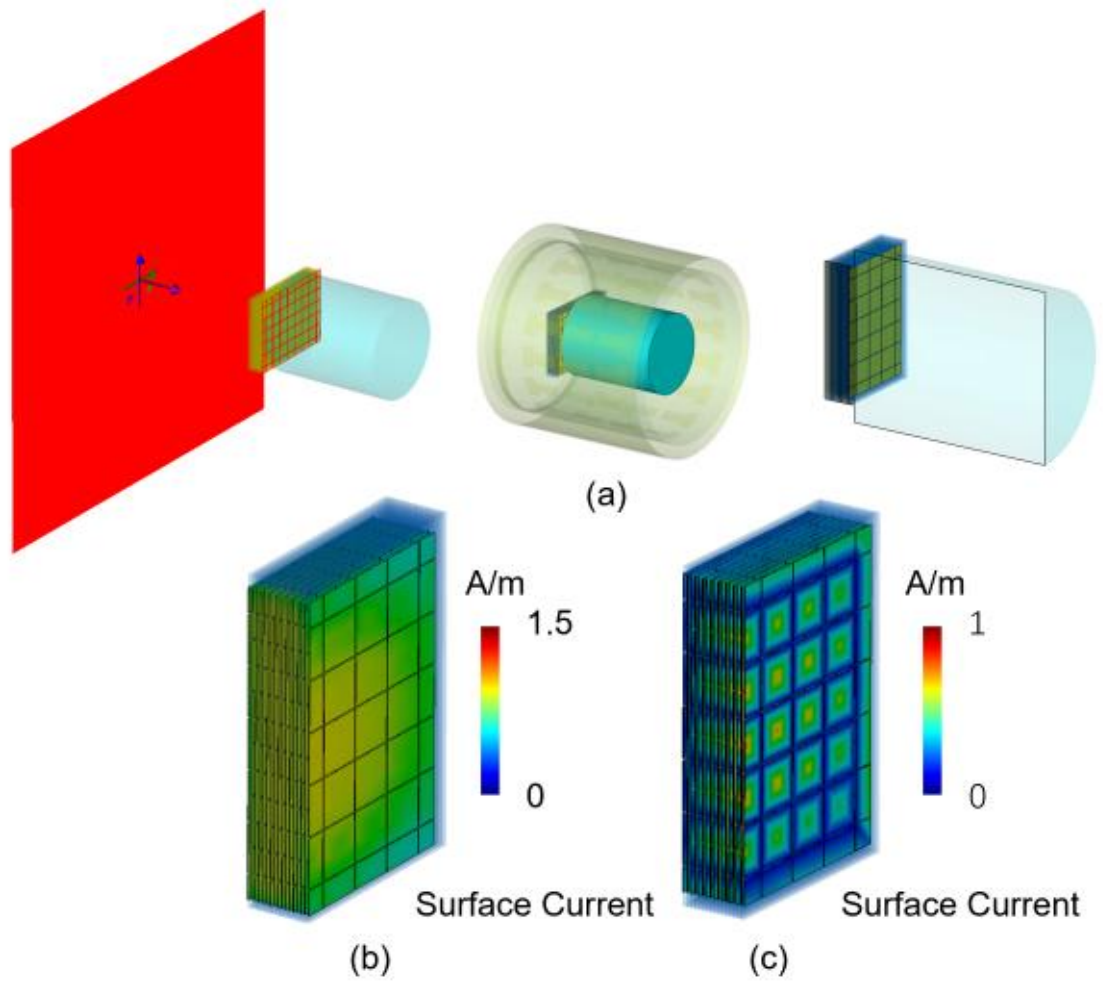

**Fig. 2. Simulation Setup and Surface Current Distribution on the ADM**: (a) Simulation setups for ADM excited by plane wave and birdcage RF field. The half-section diagram of the phantom and ADM provide detailed view of the setup. (b) Surface current distribution on the ADM under plane wave excitation, showing uniform current flow. (c) Surface current distribution on the ADM under birdcage coil's RF field excitation, demonstrating resonant current loops in each unit cell. This figure contrasts the surface current patterns induced by plane wave and reactive-region excitations, underscoring the impact of near-field conditions on ADM behaviour.

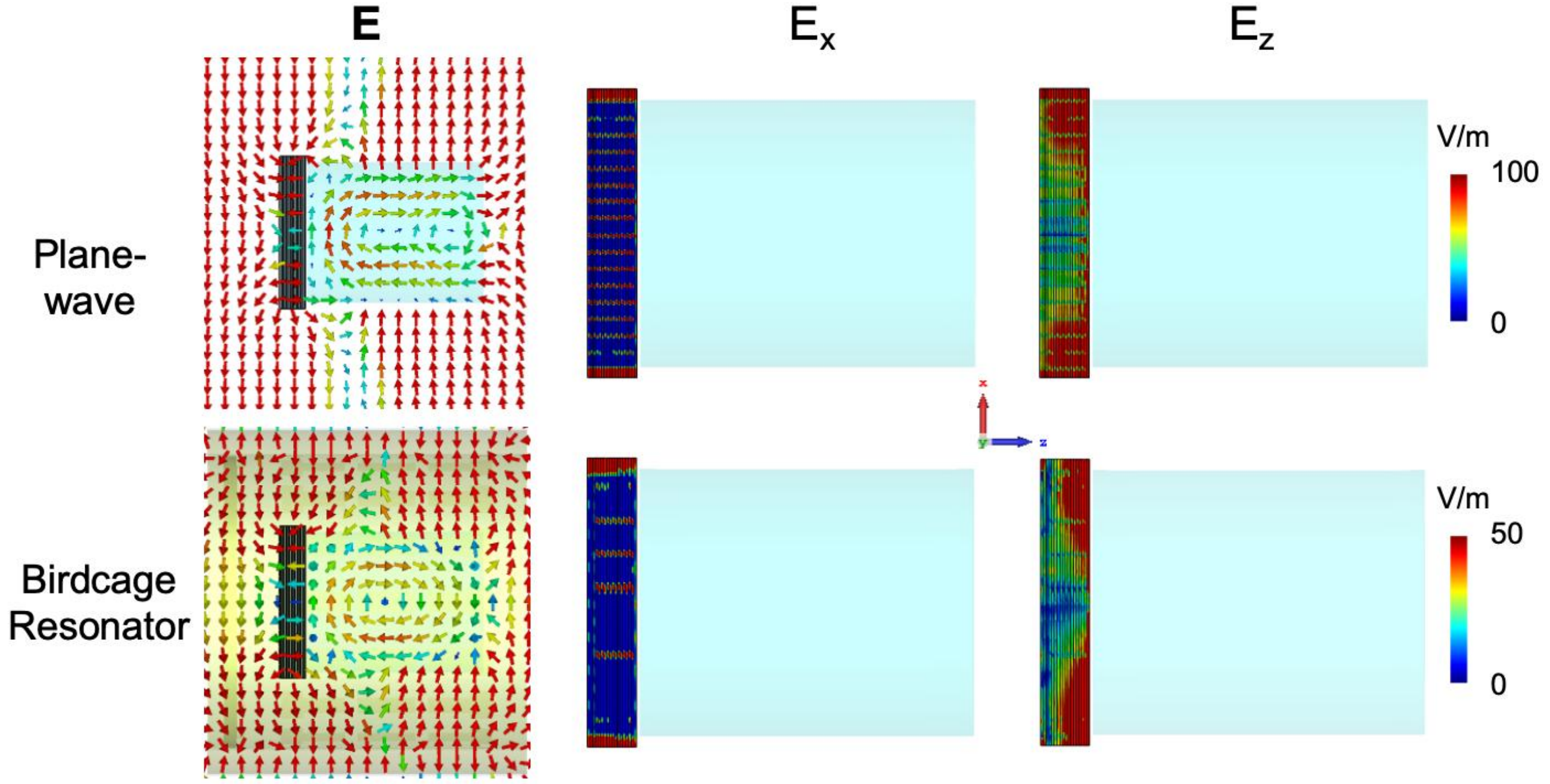

**Fig. 3. Electric-field distributions under plane-wave excitation (top row) and birdcage-coil excitation (bottom row).** Columns show the electric-field vectors, $E_x$, and $E_z$, respectively.

Full-wave simulations were performed in CST (Dassault Systèmes, France). Three substrate configurations were compared: the reference dielectric pad, an ADM designed using plane-wave-based effective parameters (ADM#1), and an ADM tuned using the proposed MSL approach (ADM#2), as shown in Figs. 4(a)–4(c). The same MSL geometry, ports, and termination were used for all three configurations.

The MSL dimensions were first selected so that the reference dielectric produced a reflection minimum at the target MRI frequency of 297.2 MHz. The reference dielectric was then replaced by ADM#1 without changing the remaining MSL measurement conditions. The ADM unit-cell dimensions were subsequently adjusted to align its reflection minimum with that of the reference dielectric, yielding ADM#2. The resulting S11 curves were compared to assess the agreement beyond the matching frequency.

Fig. 5 summarizes the workflow:

1. Select the reference dielectric: Choose a dielectric pad with known properties.
2. Configure the MSL: With the reference dielectric as the substrate, adjust the MSL dimensions to obtain a reflection minimum at 297.2 MHz.

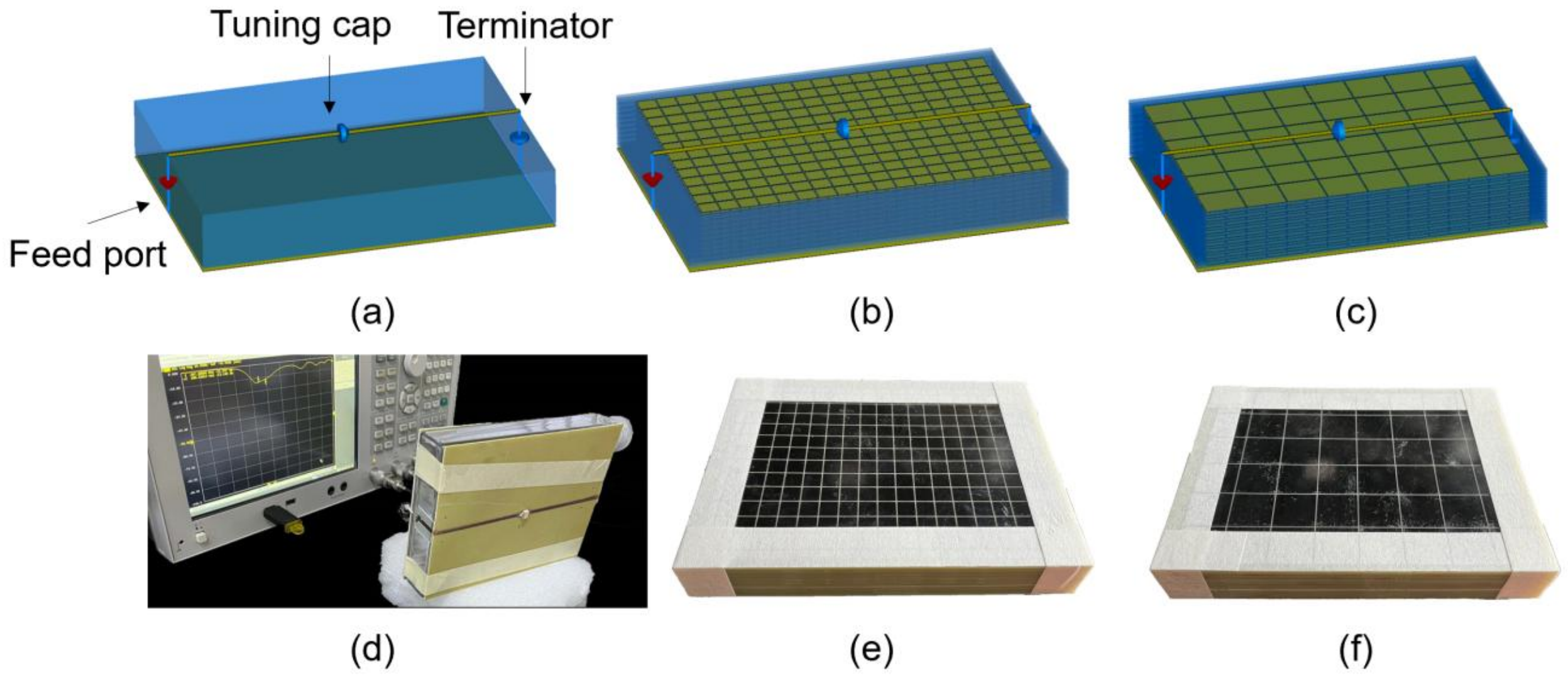

**Fig. 4. Simulation Models, Bench measurement and Fabricated ADM Pads:** (a)-(c) Simulation models of the travelling-wave microstrip-line (MSL) with different substrates: target dielectric pad, traditionally optimized ADM pad, and newly optimized ADM pad. (d) Photograph of the vector network analyzer measuring the reflection coefficient response ($S_{11}$). (e)-(f) Photographs of the fabricated traditionally optimized ADM#1 and the newly optimized ADM#2. This figure provides a comprehensive view of the simulation setup and physical construction of the ADM under test.

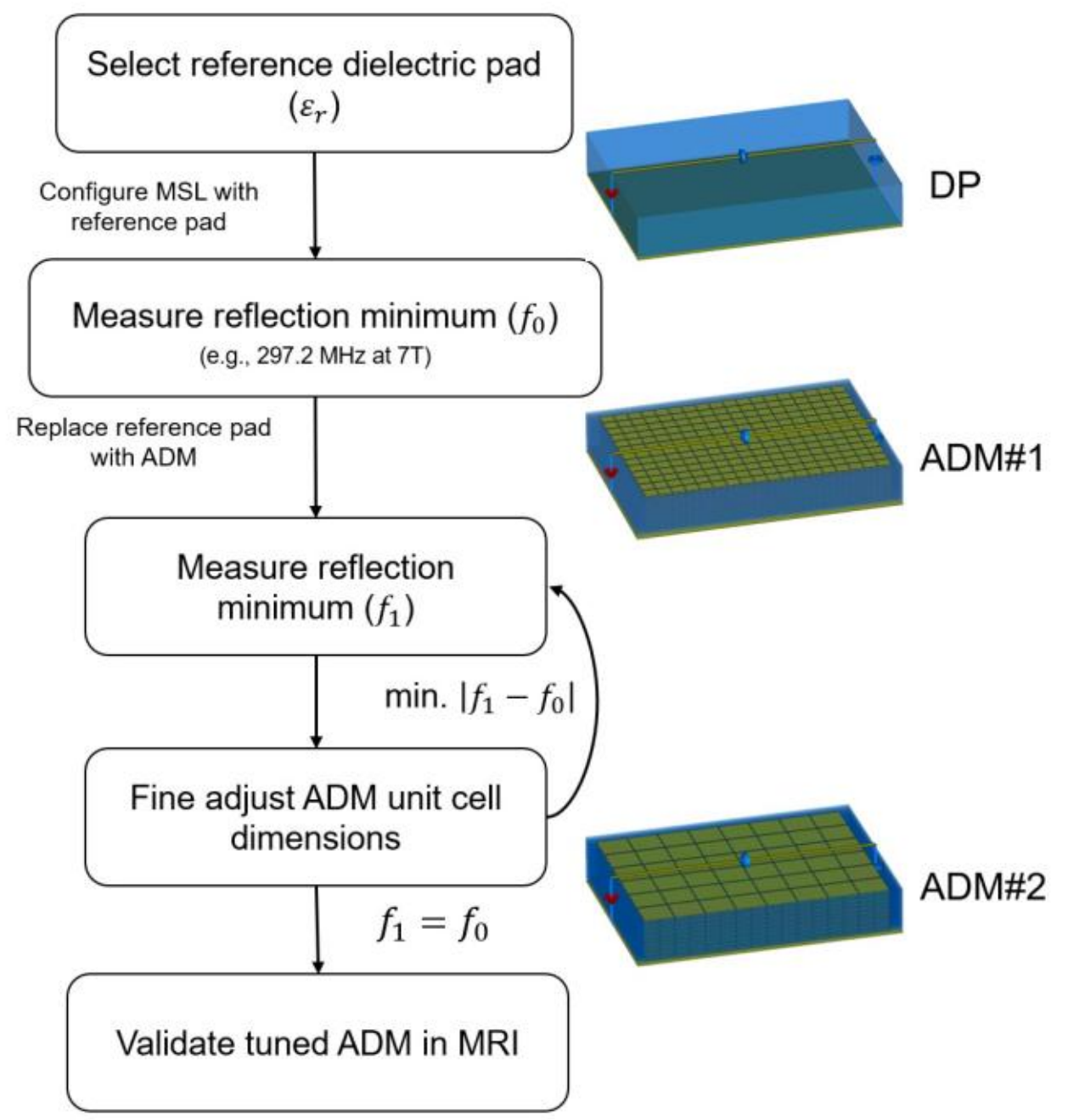

**Fig. 5. Workflow of the proposed MSL-based ADM tuning method.** The MSL is first configured with the reference dielectric pad to produce a reflection minimum at 297.2 MHz. The reference pad is then replaced by the ADM, whose unit-cell dimensions are adjusted to align the reflection-minimum frequency with that of the reference. The tuned ADM is subsequently evaluated in MRI simulations and phantom experiments.

3. Evaluate the initial ADM: Replace the reference dielectric with the plane-wave-based ADM and determine the frequency of its reflection minimum.
4. Tune the ADM: Adjust the ADM unit-cell dimensions until the reflection minimum aligns with that of the reference dielectric, while keeping the remaining MSL measurement conditions unchanged.
5. Validate in MRI: Evaluate whether the tuned ADM reproduces the reference dielectric's effect in MRI simulations and phantom experiments.

This procedure provides a practical means of matching the ADM and reference dielectric responses under the selected microstrip excitation. The applicability of this matching to the MRI environment is assessed separately, as described in the following section.

### C. Bench Measurements and MRI Validation

Bench measurements were performed using a vector network analyzer (VNA) to compare the S11 responses of the reference dielectric pad and the fabricated ADMs in the MSL test structure. The measurement setup and fabricated ADM pads are shown in Figs. 4(d)–4(f). Each material was tested using the same MSL geometry and termination conditions, allowing direct comparison of the reflection-minimum frequencies and the surrounding reflection responses. The measured responses were compared with the simulations to assess the implementation of the proposed tuning procedure.

The ADM unit-cell dimensions were further adjusted based on the bench measurements to align the reflection minimum with that of the reference dielectric at 297.2 MHz. The final geometry was then used for MRI validation.

MRI simulations were conducted using a birdcage coil operating at 297.2 MHz and a cylindrical phantom with a relative permittivity of 68 and a conductivity of 0.5 S/m. Four configurations were evaluated: no dielectric pad, the reference dielectric pad, the plane-wave-based ADM#1, and the MSL-tuned ADM#2. The simulated $B_1^+$ distributions were compared to assess how closely each ADM reproduced the effect of the reference dielectric.

The optimized ADMs were fabricated and experimentally validated on a 7 T research scanner using a birdcage coil (Nova Medical). Fig. 6 provides a visual overview of the experimental setup used for the validation of the ADMs. Fig. 6 (left) shows the saline phantom placed within the Nova coil, illustrating the baseline configuration without any additional materials. Fig. 6 (right) depicts the same setup but with the ADMs added. These images illustrate the setup of the experiments, ensuring that the conditions match those simulated and intended for MRI applications. Proton-density-weighted GRE images were acquired to assess the performance of the ADMs in comparison to the target dielectric pad. The experimental results were then compared with the simulation outcomes to verify the accuracy and effectiveness of the microstrip line method in designing ADMs for MRI applications.

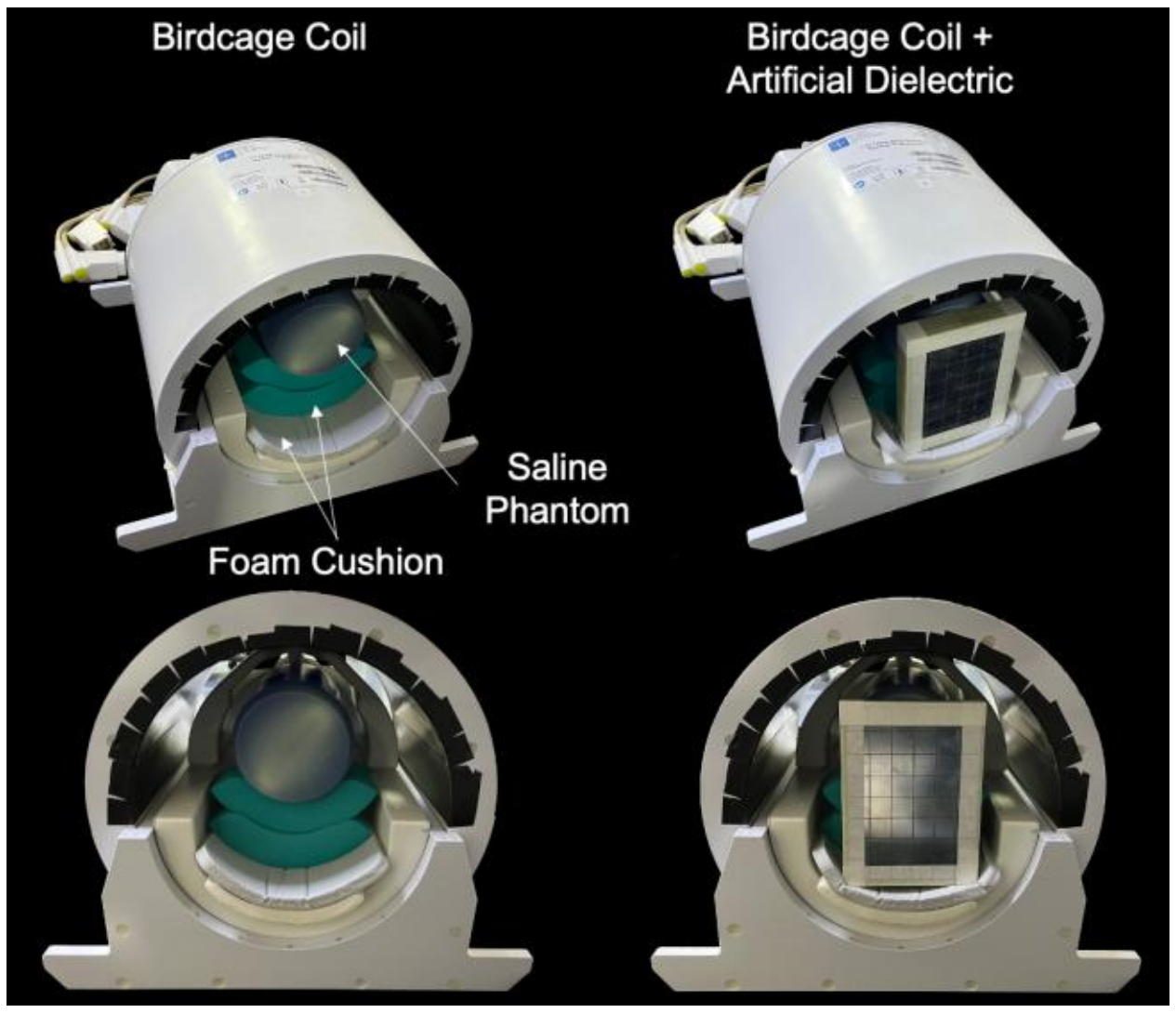

**Fig. 6. Experimental Setup** (Left) Saline phantom placed within the Nova coil without additional dielectric materials. (Right) Saline phantom with the artificial dielectric materials (ADM) integrated into the Nova coil setup.

## III. RESULTS

### A. Microstrip-Line Reflection Responses

Fig. 7 compares the simulated and measured S11 responses of the reference dielectric pad, the plane-wave-based ADM#1, and the MSL-tuned ADM#2. The reflection minima of the reference dielectric and ADM#2 occurred near the target frequency of 297.2 MHz, whereas that of ADM#1 occurred at approximately 320 MHz. The simulations and bench measurements showed the same trend, with ADM#2 more closely matching the reference dielectric's reflection-minimum frequency. These results demonstrate that the proposed tuning procedure improved the agreement between the ADM and reference dielectric responses under MSL excitation.

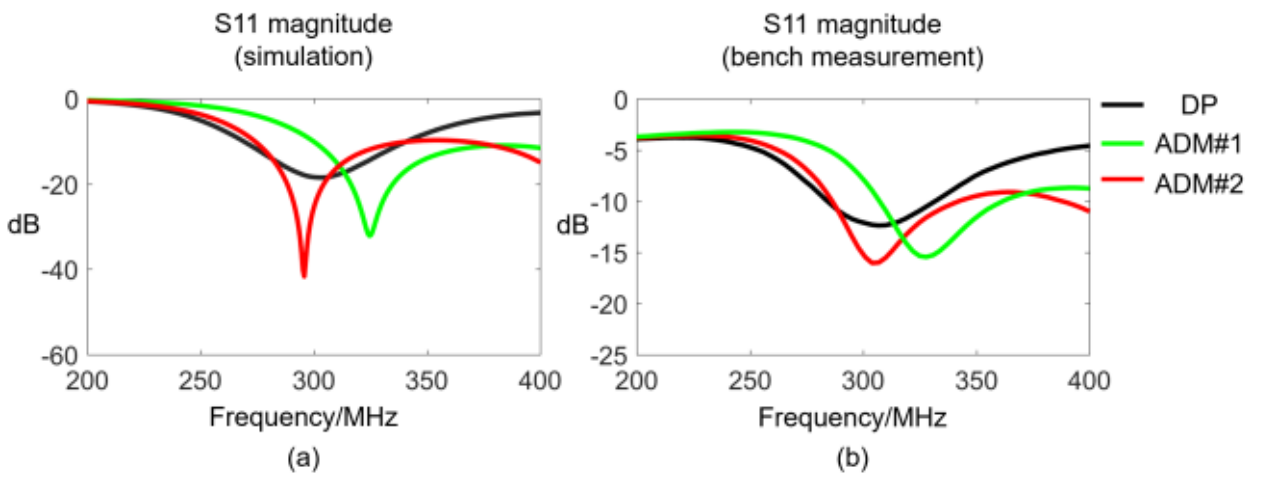

**Fig. 7. $S_{11}$ Parameter Curves from Simulation and Bench Measurement** (a) $S_{11}$ curves for the target dielectric pad, traditionally designed ADM#1, and newly designed ADM#2 from simulation. (b) Corresponding $S_{11}$ curves from bench measurements. The tuned ADM#2 more closely matches the reference pad's reflection-minimum frequency.

## *B. MRI Validation*

Fig. 8 compares four configurations: no dielectric pad, the reference dielectric pad with a relative permittivity of 78.5, ADM#1, and ADM#2. In the MRI simulations, ADM#2 produced a $B_1^+$ distribution that more closely resembled that of the reference dielectric than did ADM#1. This agreement indicates that the response matching achieved using the MSL-based method translated to a similar effect on the simulated transmit field for the configuration studied.

The experimental GRE images acquired at 7 T showed a corresponding qualitative trend. The spatial image-intensity pattern obtained with ADM#2 more closely resembled that obtained with the reference dielectric pad, whereas ADM#1 produced a visibly different pattern. These observations provide qualitative application-level support for the MSL tuning approach. Because GRE image intensity depends on both transmit and receive sensitivity, as well as acquisition parameters, the images are not interpreted as direct measurements of $B_1^+$.

# IV. Discussion

The results suggest that the MSL-based characterization provides a useful intermediate step between plane-wave-based ADM design and MRI evaluation. Although ADM#1 was designed to reproduce the effective permittivity of the reference dielectric under plane-wave excitation, its reflection minimum differed from that of the reference under MSL excitation. Adjusting the ADM geometry brought this minimum closer to the reference response, with agreement observed in both simulations and bench measurements. The tuned ADM also more closely reproduced the reference pad's simulated $B_1^+$ distribution and experimental GRE image-intensity pattern. Together, these findings support the use of MSL response matching to guide ADM tuning for the MRI configuration studied.

A possible explanation is that the spatially nonuniform excitation provided by the MSL reveals aspects of the finite ADM response that are not captured by plane-wave-based effective parameters. The current and electric-field comparisons in Figs. 2 and 3 illustrate how different excitation conditions can produce different responses within the same structure. The microstrip excitation is not identical to that of a birdcage coil, but the MRI results suggest that matching the reference response under MSL excitation can provide a useful design criterion. The extent to which this matching transfers to other coil configurations will depend on factors including field polarization, spatial field distribution, and sample geometry.

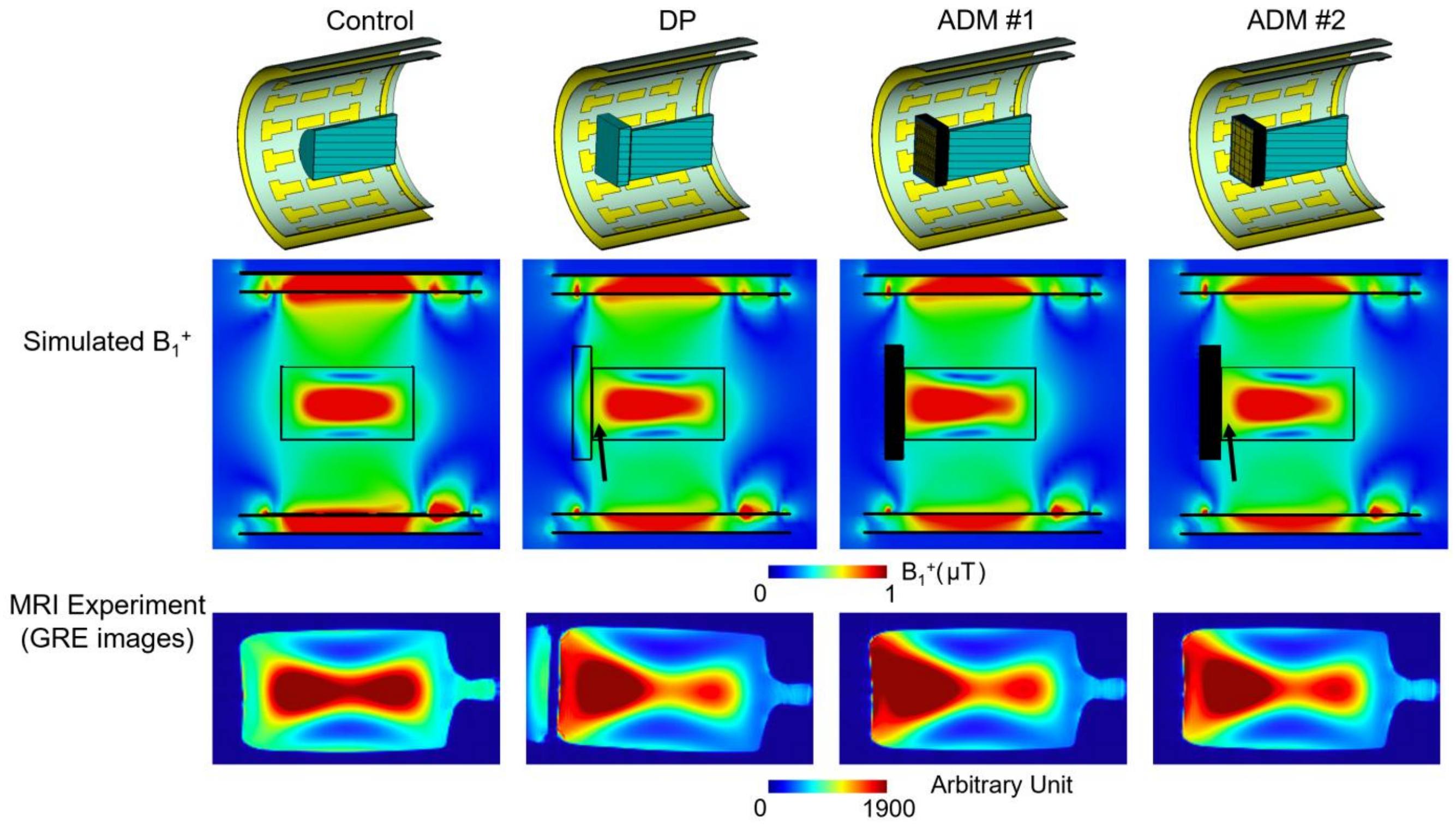

**Fig. 8. Comparison of Simulated $B_1^+$ Distributions and Experimental GRE Images:** (Upper row) Schematic diagrams illustrating the simulation setups for a cylindrical phantom within a birdcage coil under four different conditions: without any material, with a reference Dielectric Pad (DP), with an ADM optimized using traditional method, and with an ADM optimized using the proposed method. (Middle row) Simulation results showing the $B_1^+$ field distribution for each condition. (Lower row) Experimental GRE images acquired from a 7T MRI scanner, corresponding to the simulation scenarios. The figure demonstrates the $B_1^+$ field enhancement effect of the DP and the two ADMs.

The proposed approach complements established dielectric characterization techniques. Transmission/reflection methods [21], [22], resonant methods [23], open-ended coaxial probes [24], [25], and planar microwave sensors [26] characterize materials under their respective measurement conditions. Here, the objective is to tune a manufactured ADM to reproduce the response of a reference dielectric in a fixed MSL configuration. The reflection-minimum frequency provides a simple matching criterion governed by the characteristic impedance and electrical length of the loaded line. However, matching this feature alone does not uniquely establish equal bulk permittivity, dielectric loss, or response under other polarizations. The method is therefore interpreted as a comparative characterization and tuning procedure under the selected excitation conditions.

The MRI evaluation provides application-level support for this procedure. The closer agreement between the simulated $B_1^+$ distributions of ADM#2 and the reference pad indicates that the bench-based tuning was relevant to the tested coil–phantom configuration. The GRE images showed a corresponding qualitative trend, although image intensity also depends on receive sensitivity and acquisition parameters. These experiments therefore support the practical applicability of the approach without establishing quantitative equivalence of the transmit fields. Direct $B_1^+$ mapping would provide a stronger experimental assessment.

## V. Conclusion

This work presents a travelling-wave microstrip-line approach for characterizing and tuning artificial dielectric materials relative to a reference high-permittivity dielectric for MRI applications. The ADM geometry is adjusted to align its reflection-minimum frequency with that of the reference under identical MSL measurement conditions. Phantom GRE images acquired at 7 T showed qualitatively similar image-intensity patterns for the tuned ADM and reference pad. The method provides a practical bench-top step for guiding ADM design before MRI evaluation.